\documentclass[twocolumn]{aastex701}
\usepackage{amsmath}
\usepackage{amsfonts}
\usepackage{amssymb}

\usepackage{xcolor}

\usepackage{graphicx}

\newcommand{\result}[1]{#1}

\newcommand{\externalresult}[1]{#1}

\begin{document}

\title{
Can LIGO Detect Daylight Savings Time?
}


\author{Reed Essick}
\email{essick@cita.utoronto.ca}
\affiliation{Canadian Institute for Theoretical Astrophysics, University of Toronto, 60 St. George Street, Toronto, ON M5S 3H8}
\affiliation{Department of Physics, University of Toronto, 60 St. George Street, Toronto, ON M5S 1A7}
\affiliation{David A. Dunlap Department of Astronomy, University of Toronto, 50 St. George Street, Toronto, ON M5S 3H4}


\begin{abstract}
    Yes, it can.

    Catalogs produced by networks of Gravitational-wave interferometers are subject to complicated selection effects, and the gold-standard remains direct measurements of the detection probability through large injection campaigns.
    I leverage public data products from the LIGO-Virgo-KAGRA Collaborations' 3$^\mathrm{rd}$ and 4$^\mathrm{th}$ observing runs~\citep{GWTC-3-injections, GWTC-4-injections} to show that there are non-trivial temporal variations within the detection probability that are well-described by a weekly cycle.
    There are clear differences between weekends and weekdays, between day and night (at the sites), and even between daylight-savings and standard time.
    I discuss possible causes for this behavior and implications.
\end{abstract}



\section{Introduction}
\label{sec:introduction}

Ground-based Gravitational-wave (GW) interferometers, like advanced LIGO~\citep{LIGO}, Virgo~\citep{Virgo}, and KAGRA~\citep{KAGRA}, are exquisitely sensitive instruments that can detect exceptionally small differences in the lengths of their arms, typically corresponding to absolute displacements of \externalresult{$O(10^{-19})\,\mathrm{m}$} or less (see, e.g.,~\citet{Buikema:2020}).
However, the interferometers are not perfect, and an unavoidable part of interpreting astrophysical signals observed with them is understanding which types of signals are preferentially detected~\citep{Essick:2023upv}.
In the context of compact binary coalescences (CBCs), this is typically directly measured with large injection campaigns~\citep{GWTC-3-injections, GWTC-4-injections} that add simulated signals to real detector noise and process the synthetic data with real searches.
See, e.g.,~\citet{Essick:2025zed} for more discussion.

There are several well-known selections in catalogs of CBCs, such as a preference for face-on or face-off binary orientations and the fact that the detection horizon is a strong function of the component masses~\citep[see, e.g., \externalresult{Fig. 7} of][]{Essick:2025zed}.
This paper instead focuses on a more subtle selection effect: periodic changes in the detection probability throughout the week.

First identified by~\citet{Chen:2016luc} after the first observing run (O1) as a diurnal cycle, there can be a relatively strong selection of \externalresult{$\gtrsim 10\%$} in favor of detecting CBCs around midnight at the sites.
For the current detector network, this means night in North America as the two LIGO detectors in Livingston, LA and Hanford, WA dominate the overall network sensitivity.

Recently, the LIGO-Virgo-KAGRA (LVK) collaborations released the results of large injection campaigns for O3~\citep[\externalresult{April 2019 - March 2020;}][]{GWTC-3-injections} and O4a~\citep[\externalresult{May 2023 - January 2024;}][]{GWTC-4-injections}.
These publicly available datasets contain lists of injection parameters alongside significance estimates assigned by real searches.
Importantly, the injection sets were constructed so that they correspond to uniform injection rates in wall-time, regardless of whether the detectors were locked (i.e., recording science-quality data).
Therefore, effects like detector duty cycles and changes in noise properties are automatically encoded in the set of detected injections.

I focus on $p(t|\mathbb{D})$: the distribution of the time ($t$) at which a GW from a CBC passes through the center of the Earth conditioned on detection ($\mathbb{D}$; i.e., at least one search assigns the injection a \result{false alarm rate (FAR) $\leq 1/\mathrm{yr}$}).
As discussed in~\citet{Chen:2016luc}, this selection over time can naturally induce a selection across the celestial sky as the Earth rotates because GW detectors are not uniformly sensitive to all directions.
See, e.g.,~\citet{Chen:2016luc} and~\citet{Tohuvavohu:2024flg} for discussion about the implications for multi-messenger follow-up and, e.g.,~\citet{Essick:2022slj} for discussion on how this affects our ability to measure anisotropy in the astrophysical distribution of CBCs.

\begin{figure}
    \includegraphics[width=1.0\columnwidth]{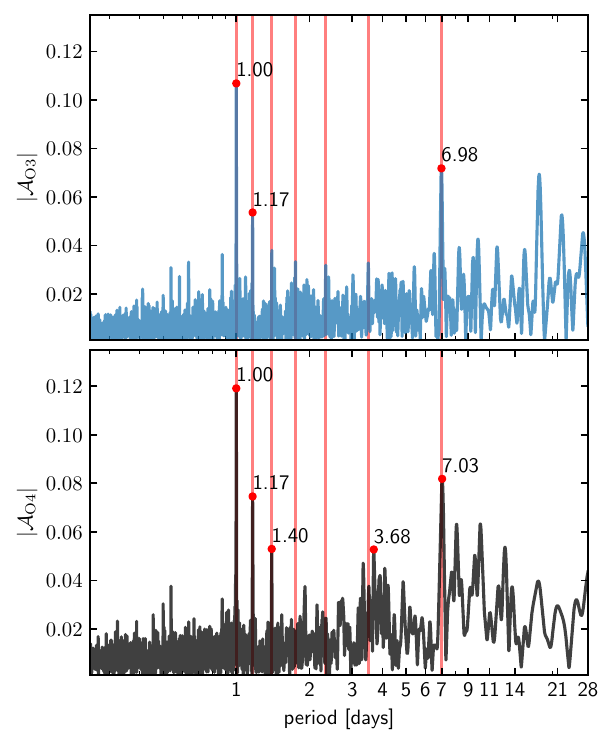}
    \caption{
        Periodograms for (\emph{top}) O3~\citep{GWTC-3-injections} and (\emph{bottom}) O4a~\citep{GWTC-4-injections} using publicly available injection data products.
        See Eq.~\ref{eq:periodogram}.
        Red lines denote periods that correspond to 1-7 cycles per week.
        Several prominent lines are annotated, including clear diurnal and weekly cycles.
        Additional periodic features are present ($|\mathcal{A}| \gtrsim 5\%$), primarily at longer periods.
        These are likely associated with long-term changes in detector sensitivity from continuous commissioning throughout the observing runs.
    }
    \label{fig:periodogram}
\end{figure}

Using the O3 and O4a injection sets~\citep{GWTC-3-injections, GWTC-4-injections}, I expand the analysis of $p(t|\mathbb{D})$.
I still find strong periodic behavior, but it is better explained by a weekly cycle rather than a diurnal cycle.
Additionally, this behavior is so clearly imprinted in modern injection datasets that it can identify the change from daylight-savings to standard time.

Sec.~\ref{sec:periodograms} identifies strong periodic features in $p(t|\mathbb{D})$ separately for O3 and O4a, showing that there is consistent behavior in both.
Sec.~\ref{sec:phase-folded histograms} then examines the behavior in more detail, identifying clear weekly trends associated with human activity.
I conclude in Sec.~\ref{sec:conclusions}.

\begin{figure*}
    \includegraphics[width=1.0\textwidth, clip=True, trim=0.0cm 1.3cm 0.0cm 0.0cm]{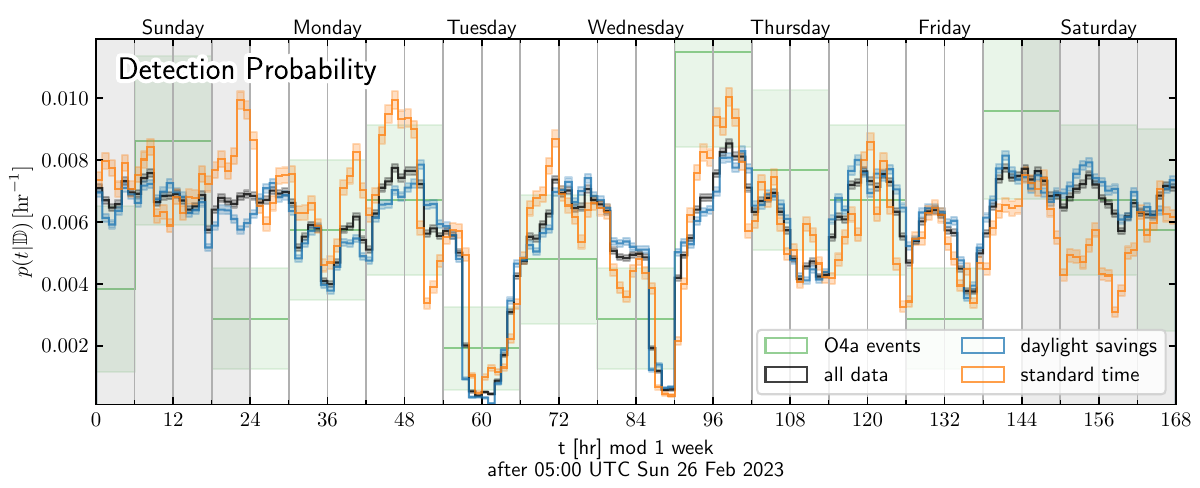}
    \includegraphics[width=1.0\textwidth, clip=True, trim=0.0cm 0.0cm 0.0cm 0.65cm]{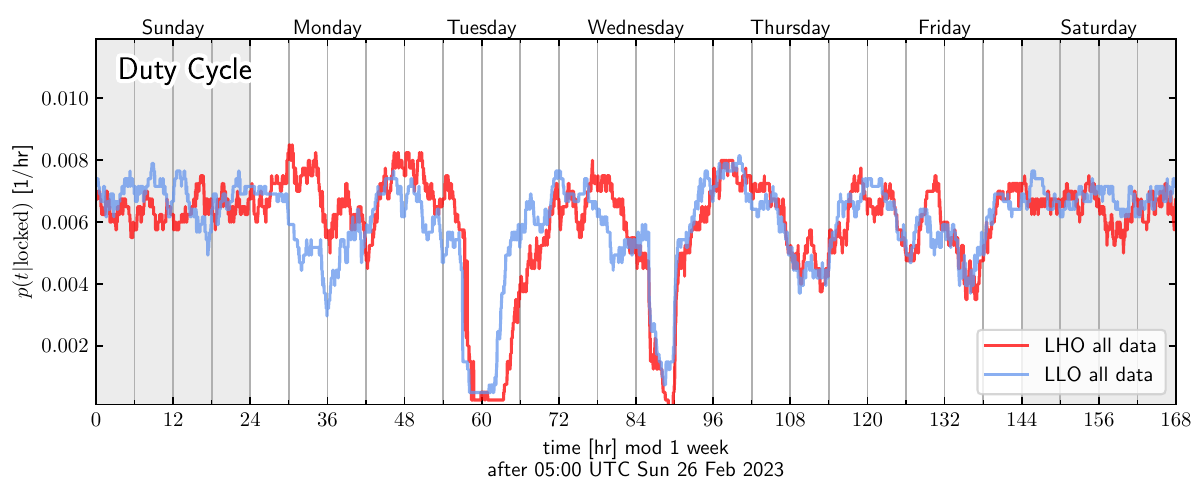}
    \caption{
        Phase-folded histogram of (\emph{top}) $p(t|\mathbb{D})$ for the times of detected injections/events from O4a and (\emph{bottom}) $p(t|\mathrm{locked})$ for the times when the detectors recorded science-quality data, both mod one week.
        Time is measured relative to midnight CDT on Sunday, and the days of the week are labeled above the axes.
        For $p(t|\mathbb{D})$, I show (\emph{black}) all the detected injections, (\emph{blue}) only injections during daylight-savings time, and (\emph{orange}) only injections during standard time.
        Each bin is one hour wide, and shaded regions correspond to 1-$\sigma$ uncertainty from the finite number of samples.
        In addition to the clear differences between weekends and weekdays (\emph{shaded and unshaded, respectively}) as well as the shift of one hour (i.e., one bin) between daylight-savings and standard time, there is complicated structure and deeper minima associated with weekly maintenance on Tuesdays and, presumably, additional emergency maintenance on Wednesdays.
        I also show (\emph{green}) the distribution of the \externalresult{87} events from O4a reported within GWTC-4.0 with \externalresult{$\mathrm{FAR} \leq 1/\mathrm{yr}$}~\citep{GWTC-4, GWTC-4-events}.
        For $p(t|\mathrm{locked})$, I show separate histograms for the LIGO Livingston (LLO) and Hanford (LHO) detectors.
        Essentially all of the weekly cycle in $p(t|\mathbb{D})$ can be accounted for by the duty cycle describing when the detectors record science-quality data.
    }
    \label{fig:folded week}
\end{figure*}


\section{Periodograms}
\label{sec:periodograms}

I begin by identifying periodic features in the set of detected events within the public injection datasets.
Because I do not have access to $p(t|\mathbb{D})$ directly, but rather only samples drawn from it, I cannot simply compute a discrete Fourier transform.
However, following the procedure introduced in~\citet{Essick:2025zed}, I can nevertheless evaluate the Fourier integral as
\begin{align}
    \mathcal{A}
        & = \int dt\, e^{-2\pi i f t} p(t|\mathbb{D}) \nonumber \\
        & \approx \sum\limits_k^N w_k e^{-2\pi i f t_k} \left/ \sum\limits_k^N w_k \right. \label{eq:periodogram}
\end{align}
where $\{t_k\}$ are $N$ samples within the injection set and the Monte Carlo sum with normalized weights $w_k$ corresponds to the expectation value with respect to the measure $p(t|\mathbb{D})$~\citep{Essick:2021}.\footnote{The weights typically correspond to the ratio of probability densities between the target distribution and the distribution used to generate the injections, including the effective rate of injections. \citet{Essick:2021} refers to the latter as ``mixture weights.''}
The absolute magnitude of Eq.~\ref{eq:periodogram} is shown in Fig.~\ref{fig:periodogram} as a function of the period separately for O3 and O4a.

Figure 8 in~\citet{Essick:2025zed} shows the same calculation for O4a and highlights that the line at \externalresult{$1\,\mathrm{day}$} is associated with the Julian day (86400 sec) rather than the sidereal day (86164.0905 sec).
This is a strong indication that these features are due to terrestrial activity.
\citet{Essick:2025zed} also identify lines at \externalresult{1.17 days}, \externalresult{1.40 days}, and \externalresult{1 week}, but do not offer an explanation for the former.
In fact, there are several clear lines that correspond to integer numbers of cycles per week, including the lines at \result{1 day (7 cycles/week)}, \result{1.17 days (6 cycles/week)}, and in O4a at \result{1.40 days (5 cycles/week)} and \result{$\sim3.5$ days (2 cycles/week)}.
This suggests that these lines are Fourier harmonics of a fundamental period of one week, and one may be better served by considering diurnal cycles as a special case that is only one part of a more general weekly cycle.

Fig.~\ref{fig:periodogram} shows that may of these lines were also present in O3 and are therefore likely caused systematically by how the detectors are operated.
Sec.~\ref{sec:phase-folded histograms} investigates the weekly cycle in detail and offers more interpretation.


\section{Phase-Folded Histograms}
\label{sec:phase-folded histograms}

\begin{figure*}
    \begin{center}
    \begin{minipage}{1.0\columnwidth}
        \includegraphics[width=1.0\textwidth]{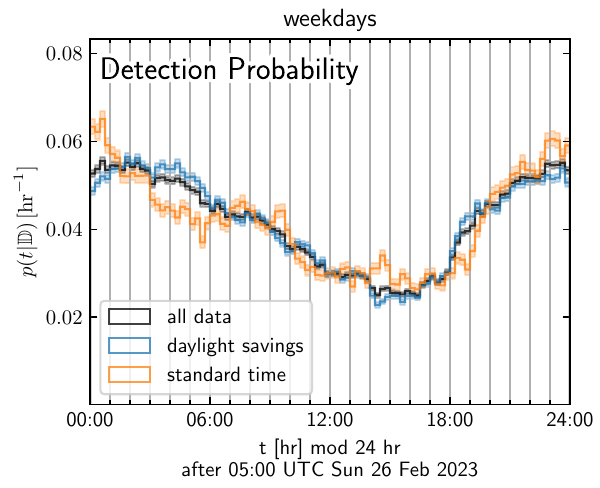}
        \includegraphics[width=1.0\textwidth]{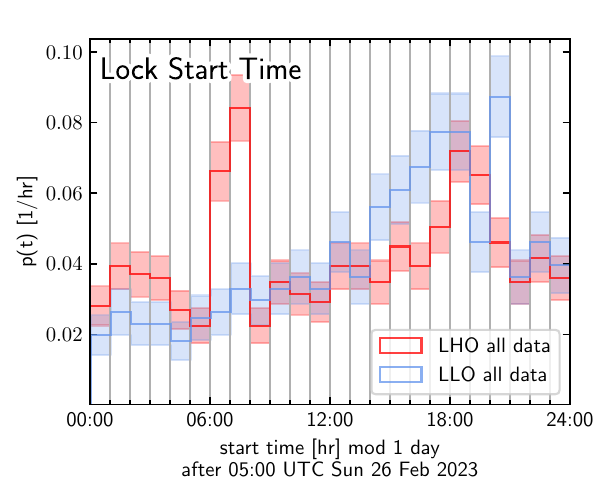}
    \end{minipage}
    \begin{minipage}{1.0\columnwidth}
        \includegraphics[width=1.0\textwidth]{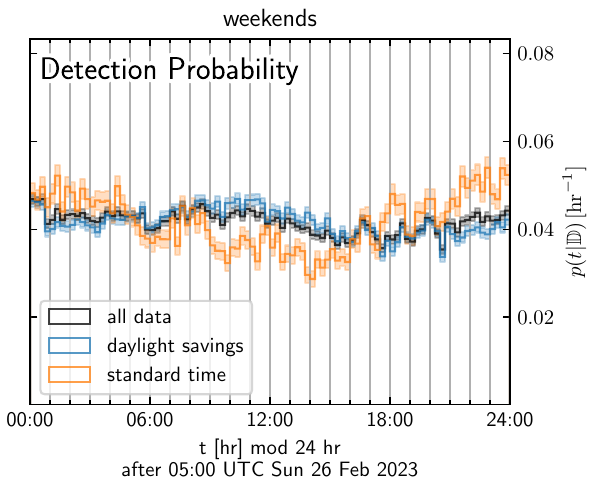}
        \includegraphics[width=1.0\textwidth]{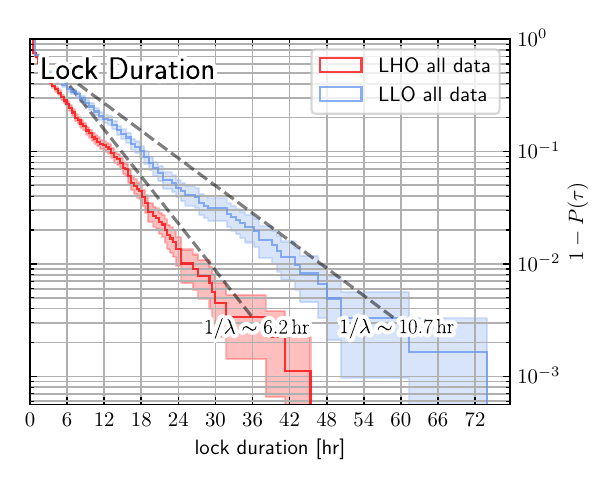}
    \end{minipage}
    \end{center}
    \caption{
        (\emph{top}) Phase-folded histograms of $p(t|\mathbb{D})$ for the times of detected events mod one day for (\emph{left}) weekdays and (\emph{right}) weekends in central daylight time (CDT).
        Each bin is 15 minutes wide; colors and shading match Fig.~\ref{fig:folded week}.
        In addition to the dramatic difference in behavior between weekdays and weekends, there is a clear shift of one hour (i.e., four bins) between daylight-savings and standard time on weekdays.
        This is particularly apparent near the sharp increase in $p(t|\mathbb{D})$ around \result{18:00 CDT}.
        (\emph{bottom left}) Phase-folded histogram for the time of day that each detector separately enters lock, showing a local maxima near \result{18:00 CDT} that corresponds to the rapid increase in $p(t|\mathbb{D})$ on weekdays.
        (\emph{bottom right}) Survival function of the lock durations ($\tau$) for each detector separately.
        These distributions are approximately exponential, corresponding to a Poisson rate of lock-loss events of \result{$\sim 1/6.2\,\mathrm{hr}$} and \result{$\sim 1/10.7\,\mathrm{hr}$} at LHO and LLO, respectively.
        There is also a small excess in LHO's distribution near \result{$\sim 12\,\mathrm{hr}$}, which matches the spacing between the two modes in LHO's lock start times.
    }
    \label{fig:folded day}
\end{figure*}

Starting with the realization that $p(t|\mathbb{D})$ is described by a weekly cycle, Fig.~\ref{fig:folded week} shows a histogram of the times of detected events relative to the days of the week in central daylight time (CDT = UTC - 5).
Events' times are folded by one week to make long-term patterns throughout O4a more apparent, and the differences between weekends and weekdays are further highlighted with shading.

Several patterns are immediately apparent.
First, there is a regular oscillation in $p(t|\mathbb{D})$ on a daily cycle for weekdays.
As expected, $p(t|\mathbb{D})$ is larger during the local night at the sites, and the roughly sinusoidal pattern had an amplitude of \result{$\sim30\%$} on Mondays, Thursdays, and Fridays during O4a.
This is similar to the cycle reported in~\citet{Chen:2016luc} for O1 and what was observed in O3 (Fig.~\ref{fig:folded day}).

However, this diurnal pattern is clearly broken in two ways: the oscillation mostly disappears on the weekends (Saturdays and Sundays) and the minima of $p(t|\mathbb{D})$ are much lower on Tuesdays and Wednesdays.
The former is a natural consequence of human activity.
If events are preferentially detected at night during the week, then they should also be preferentially detected over the weekends because both are outside of normal working hours.
Fig.~\ref{fig:folded day} highlights this for O4a by separately grouping events that were detected during weekdays and weekends folded over 24 hours.
Additionally, Fig.~\ref{fig:folded week} shows that the sensitivity over the weekends is typically comparable to the peak sensitivity during weekday nights, and the diurnal cycle is not as much of a ``peak'' in sensitivity during the night as a ``drop'' in sensitivity during the day.

The second break in the pattern is likely associated with regular detector maintenance, which always occurs on Tuesdays~\citep{O3-coincident-observation-strategy, O4a-coincident-observation-strategy, O4b-coincident-observation-strategy}.
Various housekeeping and commissioning activities occur during Tuesday maintenance.
During this time, it is very unlikely (but not impossible!) that science-quality data would be recorded.
The drop in $p(t|\mathbb{D})$ observed on Wednesdays may be associated with spill-over maintenance from Tuesdays.
That is, maintenance activities are not always completed on Tuesday, and the detectors may be taken offline on Wednesdays as well.

Fig.~\ref{fig:folded week} also shows the distribution of the \externalresult{87} new confident detections (\externalresult{$\mathrm{FAR} \leq 1/\mathrm{yr}$}) from O4a, and, as expected, the same patterns are apparent.
Although consistent with uncertainty from the finite number of events, there is a slight underabundance (overabundance) in the distribution of real detections around midnight on Sunday/Monday (Wednesday/Thursday).

The bottom panel of Fig.~\ref{fig:folded week} also shows when the detectors typically recorded science-quality data,\footnote{Specifically, Fig.~\ref{fig:folded week} shows times within segments defined by \result{\texttt{H1:DMT-ANALYSIS\_READY}} and \result{\texttt{L1:DMT-ANALYSIS\_READY}} for LIGO Hanford and Livingston, respectively.} and the main features observed in $p(t|\mathbb{D})$ are associated with the detectors' duty cycles rather than changes in noise properties during the week.

\begin{figure*}
    \includegraphics[width=1.0\textwidth]{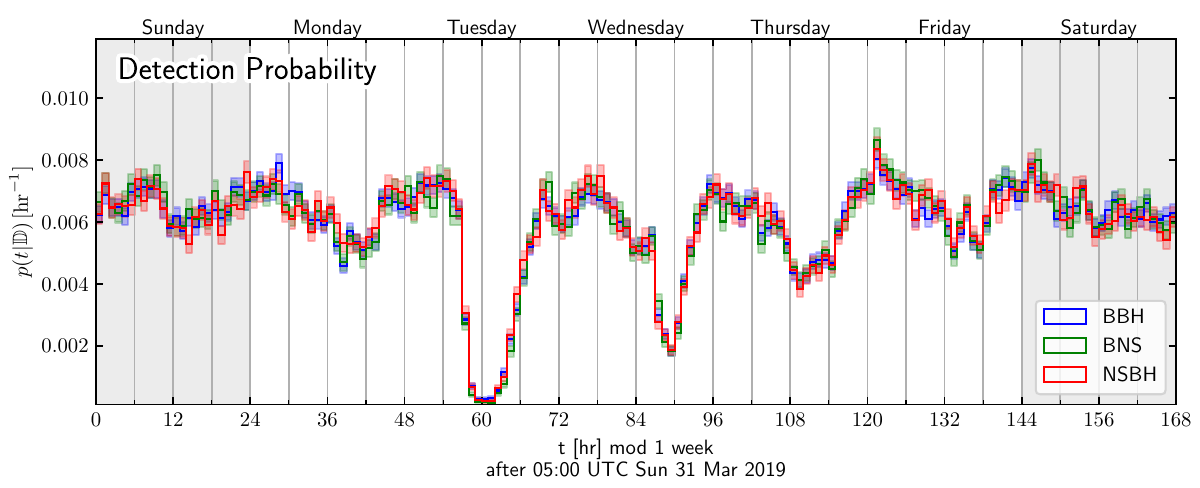}
    \caption{
        Phase-folded histogram of the times of detected events mod one week during the entirety of O3, analogous to Fig.~\ref{fig:folded week}.
        I show separate results for (\emph{dark blue}) binary black hole (BBH), (\emph{dark green}) binary neutron star (BNS), and (\emph{red}) neutron star-black hole (NSBH) systems.
        While the depth of the diurnal cycles differ between O3 and O4a, there is no evidence for differences based on component masses within O3.
    }
    \label{fig:folded week o3}
\end{figure*}

Fig.~\ref{fig:folded day} analyzes the duty cycle in more detail.
It shows that the detectors are systematically more likely to record science-quality data around \result{18:00 CDT} and that the distributions of lock durations are exponential (consistent with Poisson-distributed lock-loss events) with time constants of \result{$\sim 6.2\,\mathrm{hr}$} and \result{$\sim 10.7\,\mathrm{hr}$} for LIGO Hanford and Livingston, respectively.
This invites the interpretation that operators tend to lock the interferometers before leaving work in the evenings, and the detectors tend to record data overnight before losing lock sometime the next morning.

In general, both these patterns appear in O3 as well.
Fig.~\ref{fig:folded week o3} shows the weekly cycle for separate mass ranges: binary black hole (BBH), binary neutron star (BNS), and neutron star-black hole (NSBH) systems during O3.
The pattern is distinct from O4a, particularly in the relative depth of the troughs on Tuesdays and Wednesdays, but overall it is fairly comparable.
Additionally, there is no evidence for different behavior between different mass ranges in O3, again suggesting that these effects are associated with when the detectors record science-quality data rather than changes in the shape of the detectors' power spectral densities (i.e., changes that may preferentially affect only some masses).
Differences in the weekly patterns between O3 and O4a also explain the changes in the relative heights of different lines in Fig.~\ref{fig:periodogram}.

Finally, the switch from daylight-savings to standard time appears in both Fig.~\ref{fig:folded week} and Fig.~\ref{fig:folded day}.
Each figure shows $p(t|\mathbb{D})$ for subsets of events split into times entirely within daylight savings (CDT = UTC - 5) and entirely within standard time (CST = UTC - 6) in North America.\footnote{The switch from daylight savings occurred on 5 Nov 2023, and Figs.~\ref{fig:folded week} and~\ref{fig:folded day} split the O4a data into times within daylight savings (before 00:00 UTC 4 Nov 2023) and within standard time (after 00:00 UTC 6 Nov 2023).}
The periodic features in $p(t|\mathbb{D})$ are shifted to an hour later (relative to UTC) for events detected in CST compared to those detected in CDT.
This shift also appears in the complex phase of the Fourier coefficients of the lines in Fig.~\ref{fig:periodogram}.
For O4a, the phase shift at a period of 1 day(= 86400 sec) is \result{$\sim 18.5^\circ$ or $\sim 74$ minutes}.

More generally, seasonal trends may also exist within $p(t|\mathbb{D})$.
Indeed, it would be surprising if they were completely absent.
However, the current datasets only cover a few seasons, and, therefore, our ability to search for this behavior is limited.
Nevertheless, Fig.~\ref{fig:periodogram} hints that additional structure may exists at very long periods.
Figs.~\ref{fig:folded week} and~\ref{fig:folded day} show other differences between daylight-savings and standard time, particularly on the weekends.
This appears to be associated with a loss of sensitivity on Saturdays comparable to the rest of the weekdays during standard time.
The fraction of O4a that occurred during standard time was much less than the fraction within daylight-savings, though, and these differences may simply be due to the shorter time period.
There is comparable variability from month-to-month within daylight-savings, although without the one-hour shift observed between daylight-savings and standard time.


\section{Discussion}
\label{sec:conclusions}

As alluded to in Sec.~\ref{sec:introduction}, there are several reasons to be concerned about accurately modeling temporal variations in the detection probability beyond the intrinsically interesting imprint of human activity at the sites.
Specifically, the actual astrophysical rate of events should be constant over the duration of current observing runs.
Therefore, most population analyses may safely fix the rate of GW events to be constant in detector-frame time.
However, GW interferometers are not uniformly sensitive to signals coming from different relative directions and orientations.
This anisotropy in the detectors' response, coupled with the Earth's rotation and strong periodicity at the same period, means that variations in $p(t|\mathbb{D})$ can easily create non-trivial selections across the celestial sphere.
\citet{Essick:2025zed} demonstrate that the selection on right ascension ($\alpha$) and declination ($\delta$) caused a \externalresult{$\gtrsim 5\%$} variation in $p(\alpha, \delta|\mathbb{D})$ across the sky during O4a.

In fact, any analysis that considers anisotropy in the astrophysical distribution could be affected by variability in $p(t|\mathbb{D})$.
This includes dark siren measurements of the Hubble constant, which use galaxy catalogs to create a prior on the location of CBCs.
Incorrectly modeling the selection on $\alpha$ and $\delta$ caused by periodic variations in $p(t|\mathbb{D})$ could produce a systematic bias in which galaxies are inferred to host CBCs, although the typical angular scales over which $p(\alpha, \delta|\mathbb{D})$ varies are much larger than the angular separation between galaxies.
More general attempts to measure anisotropy in the the astrophysical distribution of CBCs would also be confounded by ignoring the weekly cycles.
This could include flexible searches for anisotropy, like what was performed in~\citet{Essick:2022slj}, as well as searches targeting specific effects like the dipolar Doppler shift caused by the solar system's proper motion~\citep{Chung:2022, Chen:2025qsl}.
Some specific models may be protected from GW selection effects because of the quadrupolar symmetry of the GW detectors' antenna patterns (see, e.g., discussion in~\citet{Essick:2022slj}).
However, these are special cases, and it is unlikely that more complicated effects would also be saved by symmetry.

Finally, as~\citet{Chen:2016luc} pointed out, there are implications for multi-messenger follow-up of GW candidates.
In addition to limits on how promptly observations can begin based on where telescopes are located on the Earth, more general scheduling algorithms may want to take into account the fact that GW candidates are less likely to be detected at certain times.
This means that observations that require long, uninterrupted integration could be scheduled on Tuesday afternoons (in North America) in order to minimize the probability that a target-of-opportunity request to follow up a GW candidate would materialize.
While this may be moot for some ground-based observatories because of their fixed relative location to the GW detectors, it could be more important for space-based telescopes.

I conclude with the reflection that the variability observed in $p(t|\mathbb{D})$, while completely reasonable \textit{post facto}, was not predicted or widely anticipated in the literature.\footnote{There are some counterexamples, such as weekly cycles previously identified in seismic activity at the sites~\citep{Daw:2004} and the rates of different classes of non-Gaussian noise artefacts~\citep[i.e., glitches;][]{Glanzer:2023}.}
Although this type of behavior can now be measured and could be modeled within semianalytic sensitivity estimates~\citep[e.g.,][]{Essick:2023toz}, it begs the question of what other hidden selections might be present within GW observations.
What's more, increasingly subtle effects will become increasingly important as the size of GW catalogs grow and more precise measurements become possible.
It therefore seems of vital importance to continue directly measuring the GW detection probability, as this guarantees that all effects will be captured, and to retain healthy skepticism if any of the various simplifying approximations within the literature are applied outside of their intended context.


\begin{acknowledgments}

I am very thankful to Daniel E. Holz for suggesting that Fig.~\ref{fig:folded week} show the distribution of real events from O4a, to Amanda Mirna Farah for catching a typo in the phase-folded histogram labels, and to Will M. Farr for providing my LSC P\&P review.
I also thank Maya Fishbach, Thomas A. Callister, Aaron Tohuvavohu, Chad Hanna, and Daryl Haggard for helpful conversations.

I am supported by the Natural Sciences \& Engineering Research Council of Canada (NSERC) through a Discovery Grant (RGPIN-2023-03346).

I am grateful for computational resources provided by the LIGO Laboratory and supported by National Science Foundation Grants PHY-0757058 and PHY-0823459.
This material is based upon work supported by NSF's LIGO Laboratory which is a major facility fully funded by the National Science Foundation.

\end{acknowledgments}


\bibliography{biblio}

\end{document}